\documentclass[aps,12pt,preprintnumbers,amssymb,nofootinbib]{revtex4}
\usepackage{pstricks}
\usepackage{amsmath}
\usepackage{graphicx}
\usepackage{color}
\usepackage{fancyhdr}
\begin{document}
\preprint{\vbox{\hbox{NCU-HEP-k054}
\hbox{Oct 2012}}}
\vspace*{.1cm}
\title{Leptonic Flavor Violating Higgs to \mbox{\boldmath $\mu\bar{\tau} + \tau\bar{\mu}$} Decay in Supersymmetry without R Parity\footnote{Talk presented by Yifan Cheng at SUSY2012  (Aug. 13 - 18), Beijing, China.}\vspace{.1cm}}

\author{Yifan Cheng and Otto C. W. Kong}
\affiliation{Department of Physics, National Central University, Chung-li, Taiwan 32054.}

\begin{abstract}
We summarized our report on leptonic flavor violating Higgs decay into
 $\mu\bar{\tau} + \tau\bar{\mu}$ under the scheme of a generic supersymmetric 
standard model without R parity. With known experimental constraints imposed, important combinations of 
R-parity violating parameters which can give notable branching ratios are listed.
\end{abstract}
\maketitle
\vspace*{-.8cm}
\section{Lepton Flavor Violation}
In the Standard Model (SM), the lepton number of each flavor is separately conserved. However, it is 
well known from the neutrino oscillation experiments that lepton flavor 
conservation should be violated \cite{superk}. In the framework of R-parity violating (RPV) supersymmetry, the requirement of neutrino masses and mixings are easily fulfilled. Moreover, it has the advantage of a richer phenomenology. Higgs to $\mu\bar{\tau} + \tau\bar{\mu}$ decay which is forbidden in the SM is our example at
hand. These kinds of lepton flavor violating (LFV) processes, have received much attention and may give us hints to physics beyond the SM. In this talk, we present the first results
of a comprehensive study on a generic supersymmetric standard model (without R parity), 
highlighting cases of most interest. More details about the topic can be found in Refs.\cite{short,long}.
\section{The generic supersymmetric standard model (without R parity) and Higgs mass matrix}
With the superfield content of the minimal supersymmetric standard model (MSSM),
the most general renormalizable superpotential can be written as
\begin{equation}
W=\epsilon_{ab}
\left[
\mu_\alpha \hat{H}^a_u \hat{L}^b_\alpha + h^u_{ik} \hat{Q}^a_i \hat{H}^b_u \hat{U}^C_k
+ \lambda^{'}_{\alpha jk} \hat{L}^a_\alpha \hat{Q}^b_j \hat{D}^C_k 
+ \frac{1}{2}\lambda_{\alpha\beta k}\hat{L}^a_\alpha \hat{L}^b_\beta \hat{E}^C_k
\right] 
+ \frac{1}{2}\lambda^{''}_{ijk}\hat{U}^C_i \hat{D}^C_j \hat{D}^C_k 
\end{equation}
where $(a,b)$ are SU(2) indices with $\epsilon_{12}=- \epsilon_{21}=1$, $(i,j,k)$ are the 
usual family (flavor) indices, and $(\alpha ,\beta)$ are extended flavor indices going from 
0 to 3. We have four leptonic superfields $\hat{L}$, which contain the components of fermion doublet 
as $l^0$ and $l^-$, and their scalar partners as $\tilde{l}^0$ and $\tilde{l}^-$. For convenience, 
we choose a flavor basis such that only $\hat{L}_0$ bears a nonzero vacuum expectation value
and thus can be identified as $\hat{H}_d$ in the MSSM. Details of the model formulation can be found in \cite{Otto}. 

With all the RPV terms, the physical scalar states 
are now a mixture of Higgses and sleptons.
The RPV terms provide new contributions to scalar mass matrices and 
hence Higgs mass. In addition, third generation quarks and squarks 
could play an important role in 
radiative corrections to the Higgs sector, and hence should be included.
Accordingly, we implement complete one-loop corrections, 
from Ref.\cite{Ellis}, to matrix elements directly relating to Higgs bosons
(including CP-even, CP-odd and charged Higgs bosons) during our computation. 
Moreover, we 
include an estimation \cite{2loop}
of the key two-loop corrections in light Higgs related elements also \footnote{Higgs bosons mix with sleptons via RPV terms, but we can still identify 
Higgses from other sleptons due to the foreseeable smallness of RPV parameters.}. 
Note that radiative RPV corrections are typically too small to be taken into account, 
thus we study tree level RPV effects only.
\section{Calculations and Numerical results}
Among all sources which can give constraints on our 
RPV parameter setting, the one from indirect evidence of neutrino 
mass (i.e., $\sum_i m_{\nu_i}\lesssim\text{1eV}$ \cite{neutrino}) is quite crucial. 
However, since it has not been completely ruled out for neutrinos having mass 
larger than 1eV, we give some comments on branching ratios as reference under 
the condition that neutrino mass is constrained only by the solid bounds
(i.e., $m_{\nu_e}<\text{3eV}$, $m_{\nu_\mu}<\text{190keV}$ and 
$m_{\nu_\tau}<\text{18.2MeV}$ \cite{neutrinoex}) as well. 

In our numerical computation, we deal directly with mass eigenstates and put all the
tree level mass matrices into the program. The mass of the Higgs bosons (and other sparticles) needed in our analysis 
are obtained by diagonalizing corresponding mass matrices numerically.
The necessary amplitudes of tree and one-loop Feynman diagrams \footnote{During numerical computation of Feynman diagrams, \textit{LoopTools} package 
is used for the evaluation of loop integrals \cite{looptools}.}
and relevant effective couplings in the model are 
derived analytically by hand.
By encoding the derived analytical formulas of the
decay amplitudes into the numerical program, values of total amplitude and
hence decay rate can be obtained. In the computation of the total decay width
of light Higgs, we include all significant decay channels   
as well as the RPV decay $h^0\rightarrow \mu\bar{\tau}+\tau\bar{\mu}$.
With the RPV partial decay width rate for the channel and total 
decay width, the branching ratio can be obtained. While the details of our parameter setting can be found in \cite{short,long}, we highlight the most interesting combinations of RPV
parameters which give significant contributions to the decay in the Table 1.

    \begin{center}
		\begin{tabular}{c@{\hspace{20pt}}c@{\hspace{20pt}}c}
		\multicolumn{3}{l}{Table 1. Interesting contributions to branching ratio of $h^0\rightarrow\mu\bar{\tau} + \tau\bar{\mu}$} \\
		\hline\hline
		RPV Parameter & With Neutrino Mass& With Relaxed\\[-10pt]
		Combinations & $\lesssim$1eV Constraint & Neutrino Mass Constraint\\
				\hline    
           $B_2 \,\mu_3\hspace{8pt} $ & $1\times10^{-15}$ & $9\times10^{-6\hspace{4pt}}$ \\ 
	         $B_3 \,\mu_2\hspace{8pt} $ & $1\times10^{-13}$ & $7\times10^{-4\hspace{4pt}}$ \\ 
           $B_1 \,\lambda_{123}$ & $1\times10^{-5\hspace{4pt}}$ & $4\times10^{-5\hspace{4pt}}$ \\
           $B_1 \,\lambda_{132}$ & $3\times10^{-5\hspace{4pt}}$ & $7\times10^{-5\hspace{4pt}}$ \\
           $B_2 \,\lambda_{232}$ & $3\times10^{-5\hspace{4pt}}$ & $6\times10^{-2\hspace{4pt}}$ \\
           $B_3 \,\lambda_{233}$ & $3\times10^{-5\hspace{4pt}}$ & $3\times10^{-2\hspace{4pt}}$  \\
           $B_2 \,A^\lambda_{232}$ & $5\times10^{-11}$ & $7\times10^{-7\hspace{4pt}}$ \\
           $B_3 \,A^\lambda_{233}$ & $5\times10^{-11}$ & $1\times10^{-7\hspace{4pt}}$   \\[2pt]

	   		\hline
		\end{tabular}\\ \end{center} \medskip\medskip

In $B_i \mu_j$ combinations, $B_3 \mu_2$ is particularly
enhanced by tau Yukawa coupling $y_{e_3}$ 
thus becoming the largest among all $B_i \mu_j$'s. On the other hand, the values of $B_i$ and $B_i \mu_j$ are highly 
constrained separately by their loop contribution to neutrino mass 
matrix \cite{Rakshit}; a non-zero $\mu_j$ will induce tree level 
neutrino mass, hence it is constrained. In the meantime, leptonic 
radiative decays like $\mu \rightarrow e\gamma$, etc. also give upper 
bounds on $B_i \mu_j$\cite{ottoleptonic}. 
 
Among all $B_i \lambda$'s, $B_1 \lambda_{123}$, $B_1 \lambda_{132}$, 
$B_2 \lambda_{232}$ and $B_3 \lambda_{233}$ are the most important
because they can contribute to the amplitudes at tree level. 
The value of $\lambda$ is constrained by charged current experiments \cite{Barbier}. 
Besides, leptonic radiative decays also give upper bounds on $B_i \lambda$ 
\cite{ottoleptonic}. We can see from the Table 1 that $B_i \lambda$ type combinations give the most important contributions among all RPV parameter combinations.
 
As to $B_i A^\lambda$ type combinations, $A^\lambda$'s do not have experimental constraints,
and can consequently take any value. In our setting ($A^\lambda$ = 2500GeV), branching ratios from $B_i A^\lambda$ can at most reach the order of $10^{-7}$. Since decay rate is proportional to amplitude square and hence $A^\lambda$ square, it is easy to see how branching ratio modifies as $A^\lambda$ increases.

From the results above, we can see that even with RPV parameters being the only source of lepton flavor violation, notable contributions to $h^0\to \mu\bar{\tau} + \tau\bar{\mu}$ with branching ratios beyond $10^{-5}$ are possible. This would lead to 
several raw events with almost no SM background in LHC with 8 TeV energy and get amplified further with the 14 TeV energy for future LHC runs. Hence, such RPV contributions on lepton flavor violation should not be overlooked in future collider experiments.


\begin{thebibliography}{99}
\bibitem{superk}
S.~Fukuda {\it et al.} [Super-Kamiokande Collaboration],
  Phys.\ Rev.\ Lett.\  {\bf 85} (2000), 3999
  [hep-ex/0009001].
M.~H.~Ahn {\it et al.} [K2K Collaboration],
  Phys.\ Rev.\ D {\bf 74} (2006), 072003
  [hep-ex/0606032].

\bibitem{short}
A.~Arhrib, Y.~Cheng, and O.~C.~W.~Kong
[arXiv:1208.4669 [hep-ph]].

\bibitem{long}
A.~Arhrib, Y.~Cheng, and O.~C.~W.~Kong
[arXiv:1210.8241 [hep-ph]].

\bibitem{Otto}
O.~C.~W.~Kong,
  Int.\ J.\ Mod.\ Phys.\ A {\bf 19} (2004), 1863
  [hep-ph/0205205].
  
\bibitem{Ellis}
M.~Carena, J.~Ellis, A.~Pilaftsis and C.~E.~M.~Wagner,
  Nucl.\ Phys.\ B {\bf 586} (2000), 92
  [hep-ph/0003180].

\bibitem{2loop}
S.~Heinemeyer, W.~Hollik and G.~Weiglein,
  Phys.\ Lett.\ B {\bf 455} (1999), 179
  [hep-ph/9903404].
  
\bibitem{neutrino}
D.~N.~Spergel {\it et al.} [WMAP Collaboration],
  Astrophys.\ J.\ Suppl.\  {\bf 148} (2003), 175
  [astro-ph/0302209].
G.~L.~Fogli, E.~Lisi, A.~Marrone and A.~Palazzo,
  Prog.\ Part.\ Nucl.\ Phys.\  {\bf 57} (2006), 742
  [hep-ph/0506083].

\bibitem{neutrinoex}
S. Eidelman {\it et al.} [Particle Data Group],
 Phys.\ Lett.\ B {\bf 592} (2004), 1.
  
\bibitem{looptools}
T.~Hahn,
  Nucl.\ Phys.\ Proc.\ Suppl.\  {\bf 89} (2000), 231
  [hep-ph/0005029].
G.~J.~van Oldenborgh and J.~A.~M.~Vermaseren,
  Z.\ Phys.\ C {\bf 46} (1990), 425.
    
\bibitem{Rakshit}
S.~Rakshit,
  Mod.\ Phys.\ Lett.\ A {\bf 19} (2004), 2239
  [hep-ph/0406168].
Y.~Grossman and S.~Rakshit,
  Phys.\ Rev.\ D {\bf 69} (2004), 093002
  [hep-ph/0311310].
S.~Davidson and M.~Losada,
  Phys.\ Rev.\ D {\bf 65} (2002), 075025
  [hep-ph/0010325].
  
\bibitem{ottoleptonic}
C.~-Y.~Chen and O.~C.~W.~Kong,
  Phys.\ Rev.\ D {\bf 79} (2009), 115013
  [arXiv:0901.3371 [hep-ph]].
  
\bibitem{Barbier}
R.~Barbier, C.~B$\acute{\text{e}}$rat, M.~Besancon, M.~Chemtob, A.~Deandrea, E.~Dudas, P.~Fayet, S.~Lavignac, G. Moreau, E. Perez and Y. Sirois,
  Phys.\ Rept.\  {\bf 420} (2005), 1
  [hep-ph/0406039].
\end{thebibliography}
\end{document}